\documentstyle[twocolumn,aps]{revtex}
\begin{document}

\draft
\title{
Bosonization of coupled electron-phonon systems}

\author{Peter Kopietz}
\address{
Institut f\H{u}r Theoretische Physik der Universit\H{a}t G\H{o}ttingen,\\
Bunsenstr.9, D-37073 G\H{o}ttingen, Germany}
\date{July 21, 1995}
\maketitle
\begin{abstract}
We calculate the single-particle Green's function of electrons
that are coupled to acoustic phonons
by means of higher dimensional bosonization.
This non-perturbative method is
{\it{not}} based on the assumption that the electronic system
is a Fermi liquid.
For isotropic three-dimensional phonons
we find that the long-range part of the Coulomb interaction
cannot destabilize the Fermi liquid state,
although for strong electron-phonon coupling the quasi-particle residue is
small.
We also show that Luttinger liquid behavior in three dimensions
can be due to quasi-one-dimensional anisotropy in the
electronic band structure {\it{or in the phonon frequencies}}.

\end{abstract}
\pacs{PACS numbers: 72.10.Di, 05.30Fk, 11.10.Ef, 71.27.+a}

\narrowtext

A famous theorem due to Migdal\cite{Migdal58} states that in
coupled electron-phonon systems
the electron-phonon vertex is for
$\sqrt{ \frac{m}{M}} \ll 1$
not renormalized by phonon corrections.
Here $m$ is the effective mass of the electrons and $M$ is the mass of the
ions.
An implicit assumption in the proof of this theorem is that
the electronic system is a Fermi liquid\cite{Fetter71}.
However, there exists experimental evidence that the normal-state properties
of the high temperature superconductors do not show typical
Fermi liquid behavior\cite{Anderson90}. Therefore it is desirable
to study electron-phonon interactions by means of non-perturbative techniques,
which do
not assume {\it{a priori}} that the system is a Fermi liquid.
Moreover, in heavy fermion systems  the parameter
$\sqrt{ \frac{m}{M}} $ is not necessarily small,
so that Migdal's theorem may not be valid. In this case
the self-consistent renormalization of the phonon energies due to the coupling
to the
electrons cannot be neglected\cite{Bardeen55,Pines89}.
In diagrammatic approaches
it is often tacitly assumed that the phonons remain
well defined collective modes\cite{Fetter71,Engelsberg63}.

In this work we shall study electron-phonon interactions by means of our
functional integral formulation of higher dimensional
bosonization\cite{Kopietz94,Kopietz95}, and show that with
this non-perturbative approach one can circumvent
the shortcomings mentioned above.
Bosonization in arbitrary dimensions has recently
been discussed by a number of
authors\cite{Kopietz94,Kopietz95,Luther79,Haldane92,Houghton93,Castro94,Frohlich95}.
As compared with more conventional operator methods\cite{Houghton93,Castro94},
the functional bosonization approach\cite{Kopietz94,Kopietz95,Frohlich95}
has the advantage that the retarded interaction between the electrons
which is mediated via the phonons can be
obtained  trivially by integrating over the phonon field at the very beginning
of the calculation.
Bosonization is most powerful
for long-range interactions that are dominated by momentum transfers
$| {\bf{q}} | \ll k_{F}$, where $k_{F}$ is the Fermi momentum.
In this case the bosonization method
developed in Ref.\cite{Kopietz94}
leads to cutoff-independent results.
The physically most relevant
Coulomb interaction  belongs to this category at
high densities, where the Thomas-Fermi screening wave-vector
$\kappa = ( 4 \pi e^2 \nu )^{\frac{1}{2}}$
is small compared with $k_{F}$.
Here $\nu$ is the density of states at the Fermi energy.
In this work we shall therefore
start from a model of electrons interacting with Coulomb forces.
For $ | {\bf{q}} | \ll k_{F}$ the
Fourier transform of the interaction  can be approximated by its
continuum limit
$f_{\bf{q}}^{cb} = \frac{ 4 \pi e^{2}}{ {\bf{q}} ^2 }$.
The screening problem will be solved explicitly by means of
our bosonization approach.
In this respect our work is complementary
to studies based on Hubbard models\cite{Kim89},
which {\it{assume}} that
the summation of all diagrams describing
screening processes effectively
leads to the replacement of the long-range
Coulomb interaction by a short-range Hubbard-$U$.

Following the classic textbook by Fetter and Walecka\cite{Fetter71},
we use the Debye model
to describe the interaction between electrons and
longitudinal acoustic (LA) phonons, and approximate
the ionic background charge
by a homogeneous elastic medium.
Although the ions in real solids form a lattice,
the discrete lattice structure is unimportant for
LA phonons with wave-vectors $| {\bf{q}}| \ll k_{F}$.
For a detailed description of this model and its physical justification
see Fetter and Walecka\cite{Fetter71}.
However, some subtleties concerning screening and phonon energy
renormalization have been ignored in Ref.\cite{Fetter71}.
To clarify these points, we first give a careful derivation of the effective
electron-electron interaction in this model via functional integration.

The dynamics of the isolated phonon
system is described via the action
 \begin{equation}
 S_{ph} \{ b \} = \beta \sum_{q} [ - i \omega_{m} + \omega_{\bf{q}} ]
b^{\dagger}_{q} b_{q}
 \; \; \; ,
 \end{equation}
where $b_{q}$ is a complex field representing the
phonons. Here $\beta$ is the inverse temperature, $\omega_{m} = 2 \pi m /
\beta$,
and $q = [ {\bf{q}} , i \omega_{m} ]$ is a collective label.
For simplicity we first
assume LA phonons with dispersion
$\omega_{\bf{q}} = c_{0} | {\bf{q}} |$.
Here $c_{0}$ is the {\it{bare}} phonon velocity,
which is determined by the {\it{short-range part}} of the Coulomb potential and
all other non-universal forces between the ions.
The strong renormalizations due to
{\it{long-range}} Coulomb forces
will be treated explicitly in this work.
As usual, the dimensionless measure for the
strength of the interaction relative to the kinetic energy
is $\nu f_{\bf{q}}^{cb} = \frac{ \kappa^2}{ {\bf{q}}^2}$.
Because $\nu f_{\bf{q}}^{cb}
\raisebox{-0.5ex}{$\; \stackrel{>}{\sim} \;$} 1$ for $| {\bf{q}} |
\raisebox{-0.5ex}{$\; \stackrel{<}{\sim} \;$} \kappa $, the Thomas-Fermi
wave-vector
$\kappa$ defines the boundary between the long-- and short-range regimes.
Note that we are implicitly assuming that the
{\it{short-wavelength}} part of the problem has already
been solved, and that the results (for example
the phonon velocity  $c_0$)
are used as input parameters
for the calculation of the effect of the long-wavelength
modes. Such a strategy is also adopted in
renormalization group approaches to
condensed matter systems\cite{Shankar94}, and is
different from the classic
self-consistent treatment of the electron-phonon system
by Bardeen and Pines\cite{Bardeen55}. Of course, the
results for physical quantities should be identical.
We shall come back to this point below in our
discussion of the phonon energy shift.

The electronic degrees of freedom are represented by a Grassmann-field
$\psi$, so that the total action of the interacting electron-phonon system is
 \begin{equation}
 S \{ \psi , b \} = S_{0} \{ \psi \} + S_{ph} \{ b \} + S_{int} \{ \psi , b \}
 \label{eq:Stotph}
 \; \; \; .
 \end{equation}
Here
 \begin{equation}
 S_{0} \{ \psi \} = \beta \sum_{k} [ - i \tilde{\omega}_{n} + \xi_{\bf{k}} ]
 \psi^{\dagger}_{k} \psi_{k}
 \end{equation}
represents non-interacting spinless electrons
with energy dispersion $\xi_{\bf{k}} = \epsilon_{\bf{k}} - \mu$,
where $\mu$ is the chemical potential.
$\tilde{\omega}_{n} = 2 \pi ( n + \frac{1}{2} ) / \beta$
is a fermionic Matsubara frequency, and $k = [ {\bf{k}} , i \tilde{\omega}_{n}
]$.
The action $S_{int} \{ \psi , b \}$ represents the
Coulomb energy associated with all charge fluctuations in the
system\cite{Fetter71},
 \begin{equation}
 S_{int} \{ \psi , b \} =
 S_{int}^{el} \{ \psi \} +
 S_{int}^{el-ph} \{ \psi , b  \} +
 S_{int}^{ph} \{ b \}
 \; \; \; ,
 \end{equation}
 \begin{eqnarray}
 S_{int}^{el} \{ \psi \} & = &
 \frac{\beta }{2 V} \sum_{q} f_{\bf{q}}^{cb} \rho_{-q} \rho_{q}
 \label{eq:Sintel}
 \; \; \; ,
 \\
 S_{int}^{el-ph} \{ \psi , b  \} & = &
 - \frac{\beta }{2 V} \sum_{q} f_{\bf{q}}^{cb} \left[ \rho_{-q} \rho_{q}^{ion}
+ \rho_{- q}^{ion}
 \rho_{q}   \right]
 \label{eq:Sintelion}
 \; \; \; ,
 \\
 S_{int}^{ph} \{ b  \} & = &
 \frac{\beta }{2 V} \sum_{q} f_{\bf{q}}^{cb} \rho^{ion}_{-q} \rho^{ion}_{q}
 \label{eq:Sintion}
 \; \; \; ,
 \end{eqnarray}
where $V$ is the volume of the system, and the Fourier coefficients of the
densities are
 \begin{eqnarray}
 \rho_{q}  & = & \sum_{k} \psi_{k}^{\dagger} \psi_{k+q}
 \; \; \; ,
 \\
 \rho_{q}^{ion}  & = & - z \sqrt{N} \frac{ | {\bf{q}} | }{\sqrt{2 M
\omega_{\bf{q}} }}
 [ b_{q} + b^{\dagger}_{-q} ]
 \; \; \; .
 \end{eqnarray}
Here $z$ is the valence of the ions, $z N$ is the
total  number of conduction electrons in the system, and it
is understood that the ${\bf{q}} = 0$-term in the sums
should be omitted due to overall charge neutrality.
At this point the following two approximations are made in
Ref.\cite{Fetter71}:
(a) the bare Coulomb interaction $f_{\bf{q}}^{cb}$ in
$S_{int}^{el-ph} \{ \psi , b \}$
is replaced {\it{by hand}} by the static screened interaction,
$ \frac{ 4 \pi e^2}{ \bf{q}^2 } \rightarrow \frac{ 4 \pi e^2}{ \kappa^2}$,
and (b) the contribution
$S_{int}^{ph} \{ b \}$ is simply dropped.
We shall see shortly that the approximation (b) amounts to
ignoring the self-consistent renormalization of the phonon
frequencies\cite{Bardeen55,Pines89}.
Although Fetter and Walecka\cite{Fetter71} argue that these approximations
correctly
describe the physics of screening, it is not quite satisfactory
that one has to rely here on words and not on calculations.
Because in our bosonization method screening can be derived
from first principles, we do not follow the
``screening by hand'' procedure of Ref.\cite{Fetter71},
and retain at this point all terms in Eqs.(\ref{eq:Sintel})-(\ref{eq:Sintion})
with the bare Coulomb interaction.

We are interested in the electronic Green's function of the interacting
many-body
system. The Matsubara Green's function can be written as a functional
integral,
 \begin{equation}
 G ( k )
 = - \beta \frac{
 \int {\cal{D}} \left\{ \psi \right\}
 {\cal{D}} \left\{ b \right\}
 e^{- {S} \{ \psi , b \} }
 \psi_{k} \psi^{\dagger}_{k}
 }
 { \int {\cal{D}} \left\{ \psi \right\}
 {\cal{D}} \left\{ b \right\}
 e^{- S \{ \psi , b \} } }
 \; \; \; .
 \label{eq:Gphdef}
 \end{equation}
Evidently the $b$-integration in Eq.(\ref{eq:Gphdef}) is Gaussian, so that it
can
be carried out {\it{exactly}}. We obtain
the following exact expression for the interacting Green's function
 \begin{equation}
 G ( k )
 = - \beta \frac{
 \int {\cal{D}} \left\{ \psi \right\}
 e^{- {S}_{eff} \{ \psi \} }
 \psi_{k} \psi^{\dagger}_{k}
 }
 { \int {\cal{D}} \left\{ \psi \right\}
 e^{- S_{eff} \{ \psi  \} } }
 \; \; \; ,
 \label{eq:Gph2}
 \end{equation}
with
 \begin{eqnarray}
 S_{eff} \{ \psi \} & = & S_{0} \{ \psi \} +
 S_{int}^{el} \{ \psi \}
 \nonumber
 \\
 & - & \beta \sum_{q}
 \left[
 \frac{  \omega_{\bf{q}} G_{\bf{q}}^2 }{ \omega_{m}^2 + \omega_{\bf{q}}^2 +
  \omega_{\bf{q}} W_{\bf{q}} } \right]
 \rho_{-q} \rho_{q}
 \; \; \; ,
 \label{eq:Seffdefph}
 \end{eqnarray}
with
 \begin{eqnarray}
 W_{\bf{q}}  & =   &
 \left[ \frac{z^2 N}{V}  \frac{ {\bf{q}}^2 }{M }  \right]
 \frac{ f_{\bf{q}}^{cb}}{  \omega_{\bf{q}} }
 \; \; \; ,
 \\
 G_{\bf{q}}  & =  &
 \left[ \frac{ z^2 N}{V}  \frac{ {\bf{q}}^2 }{ M }  \right]^{\frac{1}{2}}
 \frac{ f_{\bf{q}}^{cb}}{  \sqrt{ 2 V  \omega_{\bf{q}} } }
 \; \; \; .
 \end{eqnarray}
The last term in Eq.(\ref{eq:Seffdefph}) is the effective interaction between
the electrons mediated by the
phonons. Combining this term
with $S_{int}^{el} \{ \psi \} $,
we arrive at
 \begin{equation}
 S_{eff} \{ \psi \}
 = S_{0} \{ \psi \} + \frac{\beta}{2 V}
 \sum_{q} f_{q}
 \rho_{-q} \rho_{q}
 \; \; \; ,
 \label{eq:Seffph2}
 \end{equation}
with the total effective interaction given by
 \begin{equation}
  f_{q}
 =
 \frac{f_{\bf{q}}^{cb}}
 { 1 +  \nu f_{\bf{q}}^{cb} \frac{ \lambda  \omega_{\bf{q}}^2 }
 { \omega_{m}^2 +  \omega_{\bf{q}}^2 } }
 \; \; \; ,
 \label{eq:fefftotdef}
 \end{equation}
where the
dimensionless measure
for the strength
of the electron-phonon coupling is
$\lambda = \nu \gamma^2$, with
$\gamma^2 =
\frac{ z^2 N }{VM \nu^2 c_{0}^2}$.
It is instructive to compare Eq.(\ref{eq:fefftotdef}) with the
expression that would result from the
``screening by hand'' procedure\cite{Fetter71} described above.
The approximation (a) amounts to setting
$ G_{\bf{q}}^2 \rightarrow  \frac{ \gamma^2}{2V}
 \omega_{\bf{q}} $
in Eq.(\ref{eq:Seffdefph}), while (b)
corresponds to the replacement
$W_{\bf{q}} \rightarrow 0$. Then one obtains from Eq.(\ref{eq:Seffdefph})
the usual result
 $f_{q} \rightarrow f_{\bf{q}}^{cb} - \gamma^2 \frac{    \omega_{\bf{q}}^2 }
 { \omega_{m}^2 +   \omega_{\bf{q}}^2 }$.
Evidently the second term can be obtained by
expanding
Eq.(\ref{eq:fefftotdef}) to {\it{first order}} in $\lambda $ and
replacing $f_{\bf{q}}^{cb} \rightarrow \frac{1}{\nu}$ in the phonon term.
By performing these replacements,
one implicitly  ignores the renormalization of the
phonon energies due to the coupling to the electrons.
On the other hand, the effective interaction in Eq.(\ref{eq:fefftotdef}) is an
{\it{exact}}
consequence of the microscopic model defined above.
We shall see shortly that phonon energy shift and damping can be {\it{derived}}
from
this expression.

Because the phonons simply modify the effective density-density interaction,
we can obtain a non-perturbative expression for the
Green's function by substituting the
interaction given in
Eq.(\ref{eq:fefftotdef}) into our general bosonization formula
for the single-particle Green's function\cite{Kopietz94}.
The result for the Matsubara Green's function can be
written as
 \begin{equation}
 G (k)   =
 \sum_{\alpha} \Theta^{\alpha} ( {\bf{k}} )
 \int d {\bf{r}} \int_{0}^{\beta} d \tau
 e^{ - i [  ( {\bf{k}} - {\bf{k}}^{\alpha}) \cdot  {\bf{r}}
 - \tilde{\omega}_{n}  \tau  ] }
 G^{\alpha} ( {\bf{r}} , \tau )
 \label{eq:Galphaqtildedef}
 \;  ,
 \label{eq:Gkres2}
 \end{equation}
where $\alpha$ labels ``boxes''  that partition the degrees of freedom
close  Fermi surface, ${\bf{k}}^{\alpha}$
points to the center of box $\alpha$ on the Fermi surface, and
the cutoff function  $\Theta^{\alpha} ( {\bf{k}} )$ is unity
if ${\bf{k}}$ lies inside box $\alpha$, and vanishes otherwise.
For a more detailed description of this geometric
construction  see Refs.\cite{Kopietz94,Kopietz95,Haldane92,Houghton93}.
The interacting ``patch''  Green's function
 $G^{\alpha} ( {\bf{r}} , \tau )$
is of the form
 \begin{equation}
 G^{\alpha} ( {\bf{r}} , \tau )
 =
 G^{\alpha}_{0} ( {\bf{r}} , \tau )
 e^{Q^{\alpha} ( {\bf{r}} , \tau ) }
 \end{equation}
 where
$G^{\alpha}_{0} ( {\bf{r}} , \tau )$  is the non-interacting Green's function,
and the {\it{Debye-Waller factor}} is
 \begin{equation}
 Q^{\alpha} ( {\bf{r}} , \tau )  =
 \frac{1}{\beta {{V}}} \sum_{ q }  f^{RPA}_{q}
  \frac{ 1 -
  \cos ( {\bf{q}} \cdot  {\bf{r}}
  - {\omega}_{m}  \tau  )
 }
 {
 ( i \omega_{m} - {\bf{v}}^{\alpha} \cdot {\bf{q}} )^{2 }}
 \label{eq:DWph}
 \; \; \; .
 \end{equation}
Here $ {\bf{v}}^{\alpha} = \nabla_{\bf{k}} \xi_{\bf{k}} |_{ {\bf{k}} =
{\bf{k}}^{\alpha} }$,
and the effective interaction is
  \begin{equation}
 f^{RPA}_{q} =
  \frac{ f^{cb}_{\bf{q}} }{ 1 + f^{cb}_{\bf{q}} {\Pi}_{ph} ( q ) }
  \;  \; , \; \;
 {\Pi}_{ph} ( q ) = \Pi_{0} ( q ) +
 \frac{  \nu \lambda  \omega_{\bf{q}}^2}{ \omega_{m}^2 +  \omega_{\bf{q}}^2 }
 \label{eq:Piphdef}
 \; \; \; ,
 \end{equation}
where $\Pi_0 ( q )$ is the usual non-interacting polarization due to the
electronic degrees of freedom.
Diagrammatically Eqs.(\ref{eq:Gkres2})-(\ref{eq:Piphdef})
are the result of a controlled resummation of the entire
perturbation series, which is possible due to an underlying
Ward-identity\cite{Kopietz95,Castellani94}.
Thus, Eqs.(\ref{eq:Gkres2})-(\ref{eq:Piphdef})
contain vertex corrections to all orders in perturbation theory, {\it{including
the renormalizations of the electron-phonon vertex which are only
negligible if the Migdal theorem is valid}}.

Note that ${\Pi}_{ph}^{-1} ( q )$ can be identified with the
dressed phonon propagator\cite{Engelsberg63}, so that
it is clear that Eq.(\ref{eq:DWph}) takes into account that
the phonon dispersion changes because of the coupling to the electrons.
The renormalized phonon mode appears as
a peak in the dynamic structure factor\cite{Pines89},
 \begin{equation}
 S ( {\bf{q}} , \omega ) = \frac{1 }{\pi} Im \left\{
 \frac{ {\Pi}_{ph} ( {\bf{q}} , \omega + i 0^{+} ) }{ 1 + f_{\bf{q}}^{cb}
 {\Pi}_{ph} ( {\bf{q}} , \omega + i 0^{+} )   } \right\}
 \label{eq:Piphdyn}
 \; \; \; .
 \end{equation}
The qualitative behavior of $S ( {\bf{q}} , \omega )$
can be determined from physical considerations\cite{Pines89}.
In the absence of phonons,
 $S ( {\bf{q}} , \omega )$
consists  of a sum of two terms.
The first term $S_{col} ( {\bf{q}} , \omega )$ is due to
the collective plasmon mode.
In $d=3$ this mode
approaches  at long wave-lengths
a finite value, the plasma frequency $\omega_{pl} = \frac{ v_{F} \kappa }{
\sqrt{3}}$,
where $v_{F} $ is the Fermi velocity.
Within random-phase approximation the plasmon is not damped, so that
 $S_{col} ( {\bf{q}} , \omega ) = Z^{pl}_{ {\bf{q}}}
 \delta ( \omega - \omega_{pl} )$
with\cite{Pines89} $Z_{\bf{q}}^{pl} = \frac{\nu}{2} \omega_{pl} (
\frac{{\bf{q}} }{ \kappa} )^2$.
For $\omega \leq v_{F} | {\bf{q}} |$
the dynamic structure factor has another contribution
$S_{sp} ( {\bf{q}} , \omega )$
due to the decay of density fluctuations
into single-pair excitations, i.e. Landau damping.
In $d=3$ this is a rather featureless function.
As long as the {\it{renormalized}} phonon velocity is small
compared with $v_{F}$ and phonon damping is small, we expect
that phonons give rise to an additional narrow peak
that sticks out of the smooth background due to
$S_{sp} ( {\bf{q}} , \omega )$.  This is the dressed phonon mode.

Eqs.(\ref{eq:Piphdef}) and (\ref{eq:Piphdyn}) confirm the above picture.
It turns out that phonons are only well defined
for $ \sqrt{\lambda} \ll \frac{v_{F}}{c_{0}} $.
In the opposite limit the phonons are strongly damped and
mix with the plasmon mode.
For $ \sqrt{\lambda} \ll \frac{v_{F}}{c_{0}} $
a simple calculation shows that the phonon contribution
to the dynamic structure factor can be approximated by
 \begin{equation}
 S_{ph} ( {\bf{q}} , \omega )  \approx
   \frac{ Z_{{ \bf{q}}}^{ph} }{\pi}
 \frac{ \Gamma_{\bf{q}} }{ ( \omega - {\Omega}_{\bf{q}} )^2 + \Gamma_{\bf{q}}^2
}
 \label{eq:Srpacloseph}
 \; \; \; ,
 \end{equation}
where the renormalized phonon energy is
 \begin{equation}
 {\Omega}_{\bf{q}} =
 \omega_{\bf{q}}
 \left[ 1 +
 \frac{ \lambda }{ 1 + {\bf{q}}^2 / \kappa^2 } \right]^{\frac{1}{2}}
 \; \; \; ,
 \label{eq:phoenshift}
 \end{equation}
the phonon damping is
 \begin{equation}
 {\Gamma}_{\bf{q}} = \frac{\pi}{4} \frac{   \omega_{\bf{q}}^2  }{v_{F} |
{\bf{q}} |  }
 \frac{\lambda}{  [  1 +  {\bf{q}}^2 / \kappa^2 ]^2 }
 \; \; \; ,
 \end{equation}
and the phonon residue can be written as
 \begin{equation}
 Z_{ {\bf{q}}}^{ph}  =
 \frac{ \nu}{2} \Omega_{\bf{q}} \left( \frac{ \bf{q} }{\kappa} \right)^4
 \frac{ \lambda }{ [ 1 +  {\bf{q}}^2 / \kappa^2  ]
 [  1 +  \lambda + {\bf{q}}^2  / \kappa^2  ] }
 \label{eq:Zqph}
 \; \; \; .
 \end{equation}
Note  that at length scales small compared with $ \kappa^{-1}$
the phonon dispersion is not renormalized, while
at distances $ | {\bf{q}} |^{-1} \gg \kappa^{-1}$
density fluctuations are surrounded by a screening cloud
which modifies the phonon velocity and leads to phonon damping.
In the long-wavelength limit Eq.(\ref{eq:phoenshift})
implies that the phonon velocity is renormalized according to
 \begin{equation}
 \tilde{c} \approx c_0 \sqrt{1 + {\lambda}}
 \; \; \; .
 \label{eq:BohmStarv0}
 \end{equation}
For a spherical three-dimensional Fermi surface
this implies
in the limit of large $\lambda$
 \begin{equation}
 {c} \approx c_{0} \sqrt{ \lambda} = \sqrt{ \frac{z}{3} \frac{m}{M}}
 v_F
 \; \; \; .
 \label{eq:BohmStarv}
 \end{equation}
This result has first been derived by
Bohm and Staver\cite{Ashcroft76,Bohm50}
by means of a different self-consistent treatment of the
coupled electron-phonon system.
Note that the renormalized phonon velocity in Eq.(\ref{eq:BohmStarv})
is independent of the bare velocity $c_0$.

In order to investigate whether the system is a Fermi liquid, it is
sufficient to calculate the quasi-particle residue $Z^{\alpha}$.
The evaluation of the full momentum- and frequency dependent  Green's function
from Eqs.(\ref{eq:Gkres2})-(\ref{eq:Piphdef})
is a formidable mathematical problem that can perhaps only be
solved numerically.
The quasi-particle residue associated with patch $\alpha$ on the
Fermi surface is\cite{Kopietz94}
$Z^{\alpha} = e^{R^{\alpha}}$, where $R^{\alpha}$
is the constant part
of the Debye-Waller in Eq.(\ref{eq:DWph}).
Expressing $f^{RPA}_{q}$
in terms of the dynamic structure factor
and performing the Matsubara sum in Eq.(\ref{eq:DWph}), we obtain
in the limit $V , \beta \rightarrow \infty$
 \begin{equation}
 R^{\alpha} = - \int \frac{ d{\bf{q}}}{ ( 2 \pi )^3} ( f^{cb}_{\bf{q}})^2
 \int_{0}^{\infty} d \omega \frac{S ( {\bf{q}} , \omega )}{ ( \omega + |
{\bf{v}}^{\alpha}
 \cdot {\bf{q}} | )^2}
 \label{eq:R2ph}
 \; \; \; .
 \end{equation}
We would like to emphasize that Eqs.(\ref{eq:Gkres2})-(\ref{eq:Piphdyn}) and
(\ref{eq:R2ph}) remain valid {\it{even if phonons are
overdamped and mix with the plasmon mode.}}
It is not difficult to see that
the integral in Eq.(\ref{eq:R2ph}) exists for arbitrary $\lambda$, so that
the Fermi liquid state is stable.
To make progress analytically, we restrict ourselves in this
work to the regime
$\sqrt{\lambda}  \ll \frac{v_{F}}{c_{0}}$, where the
phonon damping is negligible.
Then we obtain
$R^{\alpha} = R^{\alpha}_{el} + R^{\alpha}_{ph}$,
where $R^{\alpha}_{el}$ is due to the above mentioned electronic terms
in dynamic structure factor.
For a spherical Fermi surface we find
$R^{\alpha}_{el} = - \frac{r_{3}}{2} ( \frac{ \kappa }{k_{F} }  )^2$,
where $r_{3} = O(1)$ is a numerical constant\cite{Kopietz94}.
The phonon contribution $R^{\alpha}_{ph}$ is obtained
by substituting Eq.(\ref{eq:Srpacloseph}) into Eq.(\ref{eq:R2ph}).
To leading order in
$\sqrt{\lambda} \frac{c_{0}}{v_{F}} $ the Lorentzian
can be treated like a $\delta$-function, so that
 \begin{equation}
 R^{\alpha}_{ph} = - \int \frac{ d{\bf{q}}}{ ( 2 \pi )^3} ( f^{cb}_{\bf{q}})^2
 \frac{Z_{\bf{q}}^{ph}}{ ( \Omega_{\bf{q}} + | {\bf{v}}^{\alpha}
 \cdot {\bf{q}} | )^2}
 \label{eq:R3ph}
 \; \; \; .
 \end{equation}
Substituting the above expressions for
$\Omega_{\bf{q}}$ and $Z_{\bf{q}}^{ph}$ into Eq.(\ref{eq:R3ph}), the
integration can be performed analytically, with the result
 \begin{equation}
 R^{\alpha}_{ph} = - \frac{1}{4} \left( \frac{ \kappa }{ k_{F} } \right)^2
   \ln ( 1 + \lambda )
   \; \; \; .
  \end{equation}
Using
$ ( \frac{ \kappa }{k_{F} }  )^2
=  \frac{ 2 e^2}{\pi v_{F}}$, we finally
obtain for the quasi-particle residue
 \begin{equation}
Z^{\alpha} = \left[ \frac{ e^{ - {r}_{3} } }{ \sqrt{ 1 + \lambda } }
\right]^{  \frac{e^2}{\pi v_{F}}  }
\; \; \; .
\end{equation}
Note that in the limit $v_{F} \rightarrow \infty$ (corresponding
to the infinite density limit) the quasi-particle residue approaches unity.
Although our bosonization approach
is only controlled for $\kappa \ll k_{F}$,
we expect that the result remains qualitatively correct at realistic metallic
densities, where
$\frac{e^2 }{ \pi v_{F}} $ is of the order of unity.
We conclude that at long wavelengths three-dimensional LA phonons
always lead to a stable Fermi liquid, although
for strong electron-phonon coupling the quasi-particle
residue can become small.
It should be kept in mind, however, that
we have only retained the long-range part of the
Coulomb interaction, so that possible instabilities
due to processes with large momentum transfers or superconducting pairing
are not included in our formalism.

It is straightforward to
generalize our results for anisotropic systems.
For example, for strictly one-dimensional
electron dispersion the polarization in Eq.(\ref{eq:Piphdef}) is given by
$\Pi_{0} (q ) = \nu \frac{ ( v_{F} q_{x} )^2}{\omega_{m}^2 + ( v_{F} q_{x}
)^2}$.
In this case we find that Eqs.(\ref{eq:Gkres2})-(\ref{eq:Piphdef})
give rise to Luttinger liquid behavior {\it{even if the
phonon dispersion is three-dimensional}}.
In the absence of electron-phonon interactions
such a model has been studied in Ref.\cite{Kopietz94b}.
Alternatively, we may couple
one-dimensional phonons to three-dimensional electrons.
Then  we should substitute in
Eq.(\ref{eq:Piphdef}) $\omega_{\bf{q}} =  c_{0} | q_{x} | $, while
choosing for $\Pi_{0} ( q )$ the usual three-dimensional
polarization. It is easy to see that in this case
the quasi-particle residue $Z^{\alpha}$ vanishes at the two points
${\bf{k}}^{\alpha} = \pm k_{F} \hat{\bf{x}}$ on the Fermi surface,
where $\hat{\bf{x}}$ is a unit vector in the $x$-direction.
At these points $| {\bf{v}}^{\alpha} \cdot {\bf{q}} |
= v_{F} q_{x}$, so that
{\it{the $q_{x}$-integration in Eq.(\ref{eq:R3ph}) is decoupled from the
remaining phase space.}} As a consequence $R^{x}$  is logarithmically
divergent.
However, the total Debye-Waller factor
$Q^{x} ( r_{x} \hat{\bf{x}} , \tau )$ remains finite.
(We use the label $\alpha = x$ for the
patch with ${\bf{k}}^{\alpha} = k_{F} \hat{\bf{x}}$.
Because $G_{0}^{\alpha} ( {\bf{r}} , \tau )$
contains a $\delta$-function
of the components of ${\bf{r}}$ orthogonal
to ${\bf{v}}^{\alpha}$\cite{Kopietz94},
it is sufficient to set ${\bf{r}} = r_{x} \hat{\bf{x}}$ and
calculate $Q^{x} ( r_{x} \hat{\bf{x}} , \tau )$.)
At $\tau = 0$ we obtain
for $| r_{x} | \gg \kappa^{-1} $
  \begin{equation}
  Q^{x} ( r_{x} \hat{\bf{x}} , 0 ) \sim - \gamma_{ph} \ln ( \kappa | r_{x} | )
  \; \; \; ,
  \end{equation}
with
 \begin{equation}
\gamma_{ph} = \frac{e^2}{\pi v_{F}} \frac{ c_{0}}{v_{F}}
 \left[ \sqrt{ 1 + \lambda } - 1 \right]
 \; \; \; .
 \end{equation}
The logarithmic divergence implies anomalous scaling characteristic for
Luttinger liquids,
with anomalous dimension $\gamma_{ph}$.
It is also easy to calculate the
quasi-particle residue in the vicinity of the
Luttinger liquid  points $ \pm k_{F} \hat{\bf{x}}$.
A quantitative measure for the deviation from these points is the
parameter $\delta  = 1 -
\frac{ | {\bf{v}}^{\alpha} \cdot \hat{\bf{x}} |}{ | {\bf{v}}^{\alpha} | }$.
For $\delta \ll \frac{c_{0}}{v_{F}}$ we
find that the quasi-particle residue vanishes as
 \begin{equation}
 Z^{\alpha} \propto \left[ \frac{ v_{F} \delta }{c_{0} } \right]^{\gamma_{ph}}
 \; \; \; .
 \end{equation}

In summary,
we have presented a non-perturbative approach to the coupled
electron-phonon system, which is {\it{not}} based on the assumption that
the electronic system is a Fermi liquid.
We have shown that the long-range Coulomb forces mediated by
three-dimensional LA phonons
cannot destabilize the Fermi liquid.
However, anisotropic electron- or phonon dispersions
can lead to small quasi-particle residues $Z^{\alpha}$
even if the electron dispersion is three-dimensional.
Although in realistic materials the phonon energies
cannot be exactly one-dimensional on general grounds, we
know\cite{Kopietz94b} that the spectral function of
Fermi liquids with small quasi-particle residue
can exhibit characteristic Luttinger liquid features
in experimentally accessible regimes.
More generally, it is tempting to speculate that the coupling between electrons
and any well defined
quasi-one-dimensional collective mode
can lead to Luttinger liquid behavior in three-dimensional
Fermi systems. In view of the chain-like structure of some of the
high temperature superconductors, this result
might  be important for the explanation the unusual normal-state properties of
these
materials.

I would like to thank
Roland Zeyher for helping me to understand phonons,
and Kurt Sch\H{o}nhammer for discussions and collaborations.
I am also grateful to Guillermo Castilla for his advise.

\end{document}